\documentstyle[12pt]{article}
\topmargin--0cm
\oddsidemargin--1mm
\begin{document}
\renewcommand{\theequation}{\thesubsubsection.\arabic{equation}}
\renewcommand{\thesubsubsection}{\arabic{subsubsection}}
\newcommand{\pl}{\partial}
\newcommand{\be}{\begin{equation}}
\newcommand{\ee}{\end{equation}}
\newcommand{\ba}{\begin{eqnarray}}
\newcommand{\ea}{\end{eqnarray}}
\def\R{\relax{\rm I\kern-.18em R}}
\def\1{\relax{\rm 1\kern-.27em I}}
\newcommand{\Z}{Z\!\!\! Z}
\newcommand{\ph}{PS_{ph}}

\bibliographystyle{unsrt}

\begin{center}
{\LARGE Constrained systems and analytical mechanics
in spaces with torsion}

\vskip 0.3cm
Sergei V. SHABANOV
\footnote{DFG fellow; on leave from Laboratory of Theoretical
Physics, JINR, Dubna, Russia}

\vskip 0.3cm
{\em Institute for Theoretical Physics, Free University of Berlin,
Arnimallee 14, Berlin D-14195, Germany}
\end{center}

\begin{abstract}
A system with anholonomic constraints where the trajectories of 
physical degrees
of freedom are autoparallels on a manifold
equipped with a general Cartan connection  is discussed.
A variational principle for the autoparallel trajectories
is derived from the d'Alambert-Lagrange principle for
anholonomic constrained systems. A geometrical (coordinate-independent)
formulation of the variational principle is given.
Its relation to Sedov's anholonomic variational principle for dissipative
systems and to Poincar\'e's variational principle in 
anholonomic reference frames is established. A modification
of Noether's theorem due to the torsion force is studied.
A non-local action whose extrema contain the autoparallels
is proposed. The action can be made local by adding
auxiliary degrees of freedom coupled to the original variables
in a special way.  
\end{abstract}

\subsubsection{Anholonomic constrained systems}
\setcounter{equation}0

There is no need to explain how important are constrained systems
in modern physics (e.g., electrodynamics, Yang-Mills theory,
general relativity, etc).  Constraints in dynamical systems 
are usually regarded as
a part of the Euler-Lagrange equations of motion which do not
involve time derivatives of order higher than one. In other words,
both constraints and equations of motion  result from the least
action principle applied to some Lagrangian. The existence of the
Lagrangian formalism is of great importance in constrained systems
because it allows one to develop the corresponding Hamiltonian
formalism \cite{dirac} and  
canonically quantize the system \cite{dirac}.
Yet, the variational principle is a powerful technical tool
to find integrals of motion of dynamical systems via symmetries
of the Lagrangian.

The Hamiltonian (or Lagrangian) constrained systems form a relatively
small class of constrained dynamical systems. Given an "unconstrained"
system whose dynamics is governed by a  Lagrangian $L=L(v,x)$,
$v^i$ and $ x^i$ being generalized 
velocities and coordinates, respectively,
one can turn it into a constrained system by imposing supplementary
conditions $F_\alpha(v,x) =0$ (constraints) which has to
be fulfilled by the actual motion of the system. There two ways
to incorporate the constraints into a dynamical description.
First, one can simply modify the Lagrangian $L\rightarrow L
+ \lambda^\alpha F_\alpha$ with $\lambda^\alpha$ being the Lagrange
multipliers and treat the latter as independent dynamical variables.
In doing so, we are led to the Lagrangian  constrained
dynamics. The other way is to supplement the unconstrained Euler-Lagrange
equations $d/dt (\partial_v L) -\partial_x L=0$
by the constraints $F_\alpha = 0$. It is well known that if the constraints
are not integrable, the two dynamical descriptions
are not equivalent \cite{pars,arnold}. 
The non-integrable constraints are called
anholonomic, and the dynamical systems described in the latter of the
above two ways  are known as anholonomic systems. There exists
no Lagrangian or Hamiltonian formalism for anholonomic
systems, i.e. they are {\em non-Lagrangian} dynamical systems \cite{pars}.
The existence of constraints implies that the dynamical system has 
non-physical degrees of freedom, meaning that the actual motion 
of the system is determined by a less number of independent parameters than 
the number of initial values of generalized coordinates and
velocities. For holonomic systems, the motion of the physical 
degrees of freedom can be obtained by applying the conventional
variational principle to the Lagrangian reduced on the constraint
surface $L \rightarrow L\vert_{F=0}$. When applied to anholonomic
constraints, this procedure leads to wrong equations of motion
\cite{pars,arnold}.
However dynamics of physical degrees of freedom in anholonomic
constrained systems may also possess "good" properties, e.g., a covariance
under some group transformations, existence of integrals of motion
with a clear physical interpretation, etc,
that is, the properties that one always wants to see in physical systems.

An example of this kind  is provided by the autoparallel and
and geodesic motions on a manifold equipped with a general connection
$\Gamma^\mu{}_{\nu\sigma}$ compatible with metric $g_{\mu\nu},
D_\mu g_{\nu\sigma}=0$,
where $D_\mu$ is the covariant derivative.
In \cite{embed} it has been shown that the autoparallels can
be realized as the trajectories of the physical degrees of freedom
in a special anholonomic constrained system, while the geodesics
can always be regarded as the trajectories in a holonomic
constrained system.
On a manifold $M$ the autoparallels and geodesics are 
determined respectively by the following equations covariant under 
general coordinate transformations
\begin{eqnarray}
D_v v^\mu&= &\dot{v}^\mu + \Gamma^\mu{}_{\nu\sigma}v^\nu
v^\sigma = 0\ ; \label{1.1} \\
\bar{D}_v v^\mu&= &\dot{v}^\mu + \bar{\Gamma}^\mu{}_{\nu\sigma}v^\nu
v^\sigma = 0\ .\label{1.2}
\end{eqnarray}
Here $D_v $ is the covariant derivative along the velocity vector
$v^\mu$ and $\bar{\Gamma}^\mu{}_{\nu\sigma} $ are the Christoffel
symbols. Both Eqs. (\ref{1.1})
and (\ref{1.2}) determine a curve that parallel-transports
its tangent vector along itself. The curve with such a property 
is the  autoparallel
when the most general connection compatible with the metric is 
used to specify the parallel transport. The geodesic motion occurs
if the natural Riemannian connection (induced by the metric) is
chosen to define the parallel transport.
The difference of (\ref{1.1}) from (\ref{1.2}) resides
in the torsion force term. Any connection compatible with the
metric can always be represented in the form \cite{schouten}
$g_{\sigma\lambda} \Gamma^\lambda{}_{\mu\nu} = \bar{\Gamma}_{\sigma\mu\nu}
+ K_{\sigma\mu\nu}$, where $K_{\sigma\mu\nu} = S_{\sigma\mu\nu} -
S_{\mu\nu\sigma} + S_{\nu\sigma\mu}$ is called the contorsion tensor, and
$g^{\sigma\lambda}S_{\lambda\mu\nu} = S^\sigma{}_{\mu\nu} =
\frac{1}{2} (\Gamma^\sigma{}_{\mu\nu} - \Gamma^\sigma{}_{\nu\mu})$
is the torsion tensor. The deviation of the autoparallels from  the
geodesics is caused by the torsion force $
K_{\mu}{}_{\nu\sigma} v^\nu v^\sigma$.

The geodesic equation (\ref{1.2}) follows from the Hamilton
variational principle applied to the action
\begin{equation}
S = \frac{1}{2} \int dt g_{\mu\nu} v^\mu v^\nu\ ,
\label{1.3}
\end{equation}
or to its relativistic analog
\begin{equation}
S = - \int ds= -\int dt \sqrt{g_{\mu\nu}v^\mu v^\nu}\ ,
\label{1.4}
\end{equation}
with $s$ being the proper time (or length) of the trajectory
(in this case, $v^\mu = dq^\mu/ds$ and $\dot{v}^\mu = dv^\mu/ds$
in (\ref{1.2})). A particle moving
along the autoparallel trajectory is an example
of a non-Lagrangian system.  
A system of differential equations of second order is called
Lagrangian if there exists a local Lagrangian such that 
the original system is equivalent to
the Euler-Lagrange equations. One can formulate the sufficient
conditions for a given dynamical system to be Lagrangian
\cite{marc}. Even if these conditions are not fulfilled, one
can still try to find a {\em non-local} action for a given
dynamical system.
In Section 6 we construct an explicit
example of a non-local action whose extrema contain the autoparallels. 
From the geometrical
point of view Eq. (\ref{1.1}) is just as good as Eq. (\ref{1.2})
and may be regarded or postulated as an equation of motion
of a spinless particle on a manifold
\footnote{Here we do not discuss physical arguments which of 
those two trajectories should be identified with a physical trajectory
of a spinless point-like particle moving in a space with torsion. An
interested reader may find physical arguments supporting both of them,
geodesics \cite{hehl} and autoparallels \cite{pon,klbook}.}.

Here we generalize the embedding procedure 
of \cite{embed} to arbitrary spaces with curvature
and torsion. Next we make use of the representation of the
autoparallel motion as a motion of an anholonomic system to
establish a variational principle for the Lagrangians
(\ref{1.3}) and (\ref{1.4}) which leads to Eq. (\ref{1.1}).
The variational principle is derived from the well-known
variational principles for anholonomic systems, such as Gauss' principle
of least constraint, H\"older and 
d'Alambert-Lagrange principles \cite{pars, arnold,var}.
But in contrast to them, it has an advantage 
that it applies to Lagrangians {\em reduced} on
the surface of constraints.
We shall also show that the new variational principle can be given
a completely covariant (coordinate-independent) formulation on a manifold
with a general Cartan connection. For this reason we shall refer to
it as a {\em covariant} variational principle. 
Its relation to the variational
principle of Poincar\'e \cite{poincare} and to that proposed 
by Sedov for dissipative
systems \cite{sedov} is explained. Finally, we propose a 
modification of the actions (\ref{1.3}) and (\ref{1.4})
by adding new auxiliary degrees of freedom so that 
the modified actions have extrema being the autoparallels
and admit the conventional Hamiltonian formalism.

\subsubsection{Autoparallels from constrained motion}
\setcounter{equation}0

Consider a metric manifold $M$ and local coordinates $q^\mu$ on it.
Let $\Gamma^\mu{}_{\nu\sigma}$ be components of a connection
on $M$ in the coordinate basis. We denote ${\cal P}(M)$ the space
of all paths in $M$, $T_qM$ the tangent space at a point $q^\mu$,
and $TM$ the tangent bundle.
Consider an auxiliary Euclidean space $\R^n$ of the dimension greater
than that of $M$, $n>\dim M$. Cartesian coordinates in $\R^n$ are
denoted by $x^i$. In the space ${\cal P}(\R^n)$ of all paths in
$\R^n$ we define a subspace of {\em conceivable} paths (i.e., of those
allowed by constraints) as an image of ${\cal P}(M)$ in the
embedding ${\cal P}(M)\rightarrow {\cal P}(\R^n)$:
\begin{equation}
x^i(s) = \int^{s} dq^\mu \varepsilon^i_\mu(q)\ ,
\label{2.1}
\end{equation}
for any path  $q^\mu(s)$ in $M$. The embedding functions
$\varepsilon^i_\mu(q)$ are smooth on $M$. From the path embedding
(\ref{2.1}) follows the embedding of the tangent
space $T_qM$ into $\R^n$
\begin{equation}
v^i = \varepsilon^i_\mu(q) v^\mu\ , \ \ \ v^\mu\in T_qM\ ,
\label{2.2}
\end{equation}
because one can always find a curve $q^\mu(s)$ passing through
a point $q^\mu$ such that $v^\mu = dq^\mu(s)/ds$.

The space $M$ can not be embedded into $\R^n$ pointwise if the
constraints (\ref{2.2}) on the tangent space (or on the velocities
of the conceivable motion) are not integrable (anholonomic
constraints):
\begin{equation}
\oint dq^\mu \varepsilon^i_\mu(q) \neq 0\ ,
\label{2.3}
\end{equation}
for any closed path in $M$. That is, the one-forms 
$\varepsilon^i_\mu dq^\mu$ are not closed, and
there exist no mapping $M\rightarrow \R^n,
x^i = x^i(q)$, induced by (\ref{2.2}).

The scalar product in $\R^n$,\ $(w,v) = \delta_{ij} w^iv^j$, induces the
metric on $M$. For any two vectors being the images of two elements of
$T_qM$ we have
\begin{equation}
(w,v) = g_{\mu\nu} w^\mu v^\nu\ ,\ \ \ \
g_{\mu\nu} = (\varepsilon_\mu ,\varepsilon_\nu)\ ,\ \ \
v^\mu,w^\mu\in T_qM\ .
\label{2.4}
\end{equation}
Next, we determine a connection induced by the path embedding.
To do so, we require that the parallel transport in $M$ must
be compatible with the embedding (\ref{2.1}). This means the
following. Consider a curve $q^\mu(s)$ passing through a
point $q^\mu\in M$ and its image $x^i(s)$.
Take a vector field $w^\mu\in T_qM$ and parallel-transport it along
the curve $q^\mu(s)$. The resulting vector is then embedded
into $\R^n$ by means of (\ref{2.2}). Let us repeat the procedure
in the opposite way. We first embed the vector $w^\mu$ and then
parallel-transport its image along $x^i(s)$, the image of $q^\mu(s)$.
The compatibility condition implies that the vectors obtained by these
procedures may differ by a vector orthogonal to the hyperplane being
the image of $T_qM$, and this should hold for any curve and any $w^\mu$.
We remark that this condition is weaker than a similar condition
of \cite{embed} and it is sufficient to obtain the most general spaces
with curvature and torsion.

An infinitesimal variation of $w^\mu$ under the parallel transport
is proportional to $D_vw^\mu$, where $v^\mu= \dot{q}^\mu$,
while the corresponding variation of its image, $w^i =
\varepsilon^i_\mu w^\mu$, is $D_v w^i=d_vw^i = dw^i/ds$ because
the connection in $\R^n$ is trivial. The compatibility condition
means that
\begin{equation}
\frac{dw^i}{ds} - \varepsilon^i_\mu D_v w^\mu = f^i_{\mu\nu}v^\mu
w^\nu\ , \ \ \ \ (\varepsilon_\sigma,f_{\mu\nu}) =0\ .
\label{2.5}
\end{equation}
Computing the derivative in the left-hand side of (\ref{2.5})
and assuming $v^\mu$ and $w^\mu$ to be arbitrary
we find
\begin{equation}
D_\mu \varepsilon^i_\nu = f^i_{\mu\nu}\ .
\label{2.5a}
\end{equation}
Multiplying this equation by $\varepsilon^i_\sigma$ we obtain
the connection coefficients
\begin{equation}
\Gamma^\mu{}_{\nu\sigma} = g^{\mu\lambda}(\varepsilon_\lambda,
\partial_\nu\varepsilon_\sigma)\ ,
\label{2.6}
\end{equation}
where $g^{\nu\mu}g_{\mu\sigma} =\delta^\nu_\sigma$.
It is easy to verify that the compatibility condition holds
for tensors of higher rank, too, provided the induced
metric is used to lower and rise tensor indices. 
Thanks to the compatibility condition, the induced connection
turns out to be compatible with the induced metric
$D_\mu g_{\nu\sigma}= (D_\mu\varepsilon_\nu,\varepsilon_\sigma)
+(\varepsilon_\nu,D_\mu\varepsilon_\sigma)= 0$.
In the non-integrable case (\ref{2.3}),
the induced connection may have a torsion
\begin{equation}
S^\mu{}_{\nu\sigma} = \frac 12 g^{\mu\lambda}\left[
(\varepsilon_\lambda,\partial_\nu\varepsilon_\sigma)-
(\varepsilon_\lambda,\partial_\sigma\varepsilon_\nu)\right]\ .
\label{2.7}
\end{equation}
The tensor fields $f^i_{\mu\nu}$ must also satisfy the 
integrability condition for Eq. (\ref{2.5a}) resulting 
from the commutation relation 
\begin{equation}
[D_\mu,D_\nu] = R_{\mu\nu}{}^{\sigma\lambda}L_{\sigma\lambda}
- 2S^\sigma{}_{\mu\nu} D_\sigma\ ,
\label{2.7a}
\end{equation}
where $L_{\mu\nu}= -L_{\nu\mu}$ are generators of a (pseudo)orthogonal
group acting in the tangent space of $M$, and 
$R_{\mu\nu}{}^{\sigma\lambda}$ is the Riemann-Cartan curvature
tensor. Making use of the integrability conditions we derive
\begin{equation}
(f_{\mu\lambda}, f_{\nu\sigma}) - (f_{\nu\lambda},f_{\mu\sigma})
= 2R_{\mu\nu\lambda\sigma}\ .
\label{2.7b}
\end{equation}
That is, the tensor fields $f^i_{\mu\nu}$ determine the
Riemann-Cartan curvature tensor on $M$. The case $f^i_{\mu\nu}
=0$  corresponds to the so called teleparallel spaces with
zero curvature and nonzero torsion.
With sufficiently large $n$ (at least one should have $2n 
> 1 + (\dim M)^2$, the number of independent 
components of the metric and torsion tensors), 
one can construct the most general connection 
on $M$.

Let us analyze dynamics induced by the anholonomic constraints
(\ref{2.2}), assuming the unconstrained motion to be a free motion
in the Euclidean space $\R^n$.
There are several variational principles for anholonomic
systems \cite{var} that can be applied to our system
to derive the equations of motion. If the constraints
are integrable, they all are equivalent to the conventional
Hamilton variational principle. 

In the auxiliary Euclidean
space the states of the system is labeled by  pairs $\psi =(v^i,x^i)$.
For any two trajectories $x^i_{1,2}(s)$ passing through 
the state $\psi$, Gauss'
deviation function is defined as \cite{arnold,pars}
\begin{equation}
G_\psi = \frac 12\left(\dot{v}^i_1 -\dot{v}^i_2\right)
H_{ij} \left(\dot{v}^j_1 -\dot{v}^j_2\right)\ ,
\label{2.9}
\end{equation}
where $H_{ij}(\psi) = \partial^2L/\partial v^i\partial v^j$
is the Hessian for a Lagrangian $L$ in the state $\psi$.
Gauss' principle of least constraint says that the deviation
of conceivable motions (allowed by the constraints) from
the released (unconstrained) motion takes a stationary value
on the actual motion. A physical meaning of Gauss' principle
is transparent: the acceleration (or the force) caused by
the constraints must have a minimal deviation from
the acceleration of the unconstrained motion. 
In our system the released motion
is the free motion $\dot{v}^i=0, H_{ij}= \delta_{ij}$, the accelerations of
the conceivable motion are obtained by taking a time 
derivative of (\ref{2.2}). The Gauss deviation function
assumes the form 
\begin{equation}
G_\psi = \frac 12 g_{\mu\nu}D_vv^\nu D_vv^\mu + \frac 12 (f_v,f_v)\ ,\
\ \ \ \ f_v^i = f_{\mu\nu}^i v^\mu v^\nu\ .
\label{gauss}
\end{equation}
The second term in the deviation function does not depend on
the acceleration at the physical state $\psi = (q^\mu,v^\mu)$, while
the first term 
is non-negative and attains its absolute minimum if
$D_vv^\mu =0$, i.e., for the autoparallel trajectories.

Let us denote the Lagrange derivative as $[L]_i =
-d/ds(\partial_{v^i} L) +\partial_{x^i} L$.
We remind that elements of the tangent space are also
called {\em virtual velocities} in analytical mechanics.
The d'Alambert-Lagrange principle asserts that
the conceivable motion of a system with the Lagrangian $L$
is an actual motion if for every moment of time \cite{arnold,pars}
\begin{equation}
(w,[L]) =0\ 
\label{2.10}
\end{equation}
for all virtual velocities of the 
constrained motion. We take $L=(v,v)/2$ and calculate
$[L]_i$ for the conceivable motion $[L]_i =- \dot{v}^i
=-d/ds( \varepsilon^i_\mu v^\mu)$. 
For virtual velocities of the constrained motion 
$w^i=\varepsilon_\mu^i w^\mu$,
Eq. (\ref{2.10}) 
assumes the form $(w,[L]) =- g_{\mu\nu}w^\nu D_vv^\mu =0$. For an
arbitrary $w^\mu$, it leads to the autoparallel
equation (\ref{1.1}). 

There is an equivalent formulation of the d'Alambert-Lagrange
principle known as H\"older's variational principle \cite{arnold,pars}.
A conceivable path is called a critical point of the action
functional if its variation vanishes when restricted on the
subspace of virtual velocities of the constrained motion. 
H\"older's variational
principle assumes that the actual constrained motion is
the critical point of the action. For an infinitesimal
variation of the trajectory we have $\delta x^i(s) = uw^i(s), u\rightarrow 0$.
Then $\delta S[x]/\delta x^i = \partial_u S[x+\delta x]\vert_{u=0}
=0$. This is equivalent to (\ref{2.10}). Restricting the virtual
velocity $w^i$ to the subset specified 
by the constraint (\ref{2.2}) we arrive
at the autoparallel equation (\ref{1.1}).

All the conventional variational principles for anholonomic
systems are applied to {\em non-constrained} Lagrangians,
while the constraints are implemented through the restriction
of path variations to a {\em specific} class determined by the 
constraints. We shall now develop this idea further and
find an equivalent variational principle that applies to
the Lagrangian {\em restricted} on the surface of constraints.
We recall that for anholonomic constraints the 
Hamilton variational principle leads to wrong equations
of motion, when applied to the Lagrangian restricted on
the constraint surface. In particular, we shall find
a variational principle for
the Lagrangian (\ref{1.3}) or (\ref{1.4}) that leads to
the autoparallel equation (\ref{1.1}).

Before we proceed let us make a remark. Our treatment of the
autoparallel motion as an anholonomically constrained motion
is somewhat unconventional. Typically, anholonomic 
constraints are imposed
directly on positions and velocities of the unconstrained
motion, i.e. on $x^i$ and $v^i$ (e.g., a skater on an inclined 
plane, a rolling ball on a rough surface, etc, 
\cite{arnold,kozlov}). They can be 
regarded as restrictions on the {\em initial} values of velocities
and positions so that solutions of equations of motion depend
on the less number of parameters. 
A historical reason for such a treatment
is that the positions $x^i$ can be measured and have a natural
physical meaning in many concrete
anholonomic systems \cite{kozlov}. 
It is important to observe that 
for anholonomic constraints there exists no subspace in the 
original configuration space that could be identified with
the physical configuration space formed by initial values
of {\em positions} allowed by the constraints.
In our anholonomic system, we have explicitly introduced
a physical configuration space $M$, while the auxiliary
Euclidean space is used only to formulate a dynamical principle
that specifies the  motion in $M$. 
Since anholonomic constraints do not allow us to regard
$M$ as a submanifold in the configuration space $\R^n$ of the
unconstrained system,
we have imposed constraints by restricting paths in $\R^n$ 
to a subclass permitted by the constraints, i.e.,
through the embedding of all trajectories of physical degrees of freedom 
${\cal P}(M)$ into ${\cal P}(\R^n)$. 
This is always possible for any type of constraints.

\subsubsection{Variational principle on manifolds}
\setcounter{equation}0
  
Here we shall give a formulation of the variational
principle on manifolds that is convenient for the
subsequent generalization to anholonomic systems.

Consider a vector field $w^\mu(q)$ on $M$ satisfying the
boundary condition
\begin{equation}
w^\mu(q_1) = w^\mu(q_2) =0\ ,
\label{3.1}
\end{equation}
and an action functional $S=\int ds L(v,q)$ for any
trajectory connecting the points $q^\mu_{1,2}$.
We {\em define} a variation of the action relative the vector
field $w^\mu$ by
\begin{equation}
\delta_w S = \int ds \left(\frac{\partial L}{\partial v^\mu}
d_wv^\mu + \frac {\partial L}{\partial q^\mu}w^\mu\right)\ .
\label{3.2}
\end{equation}
The derivative $d_wv^\mu$ of the velocity along the variation vector 
field specifies the variation of $v^\mu$. Given a trajectory $q^\mu
= q^\mu(s)$, the velocity vector field $v^\mu$ is known only
along the trajectory, while we need to know the behavior of $v^\mu$
in the {\em vicinity} of the trajectory in order to calculate
$d_wv^\mu$ for a generic $w^\mu$. We require 
\begin{equation}
{\cal L}_w v^\mu \equiv d_wv^\mu - d_vw^\mu\ =0\ ,
\label{3.3} 
\end{equation}
where ${\cal L}_w v^\mu =- {\cal L}_v w^\mu$ denotes
the Lie derivative on $M$. Thus, for any Lagrangian,
being a function on the tangent bundle $TM$, the smooth
vector field $w^\mu$ on $M$ determines a variation 
of the position, and Eq. (\ref{3.3}) specifies the
variation of the second independent coordinate on $TM$,
the velocity $v^\mu$. Eq. (\ref{3.3}) can be given 
a tensor form symmetrical relative the velocity and variation vector fields
\begin{equation}
{\cal L}_w v^\mu = {\cal L}_v w^\mu\ .
\label{3.3a}
\end{equation}
The Euler-Lagrange equation
follows from $d_v w^\mu = dw^\mu/ds$, the boundary condition (\ref{3.1})
and $\delta_w S =0$
that should hold for any $w^\mu$. 
Since the Lie derivative of a tensor is
a tensor \cite{schouten}, Eqs. (\ref{3.3}) and (\ref{3.3a}) are covariant
under general coordinate transformations, and so are the corresponding
equations of motion if the Lagrangian is a scalar.   

The above geometrical formulation of the variational principle 
is equivalent to the conventional one where
the variation of the action is defined via smooth path variations. Indeed,
equation (\ref{3.3}) implies that for any trajectory $q^\mu(s)$
with fixed endpoints there exists a one-parameter family of 
trajectories $q^\mu(s,u)\equiv q_u^\mu(s)$ with the same endpoints such that
$q^\mu(s,u=0) = q^\mu(s)$ and $w^\mu = \partial_u
q^\mu(s,u)$ in the vicinity of the trajectory $q^\mu(s)$. 
In other words, there exists a local coordinate net $q^\mu(s,u)$
such that $v^\mu$ and $w^\mu$ are tangent vectors for the coordinate
lines $u=const$ and $s=const$, respectively. The variable $u$ plays the
role of the variation parameter so that  
$\delta_w S[q] = d/du S[q_u]\vert_{u=0}$.

If the velocity variation is specified by means of a new
principle, other than a smooth deformation of the path, 
the condition (\ref{3.3}) would get modified and, therefore,
the variational principle would yield new equations of motion.
To obtain such a new principle for the autoparallel motion, we
make use of the variational principles
for anholonomic systems discussed in Section 2. If the condition
(\ref{3.3}) is dropped, then there exist no coordinate net
$q^\mu(s,u)$ such that $w^\mu = \partial_u q^\mu(s,u)$ if
$v^\mu = \partial_s q^\mu(s,u)$. For this reason the variation
(\ref{3.2}) will be  called non-coordinate or anholonomic. Note
that $w^\mu$ remains a smooth vector field on $M$. 

\subsubsection{Covariant variational principle}
\setcounter{equation}0

On a path 
$q^\mu(s)$ with fixed endpoints, consider a vector 
field $w^\mu$ that satisfies the boundary condition (\ref{3.1}).
Its image $w^i(s) =\varepsilon^i_\mu(q(s))w^\mu(q(s))$ determines
a variation vector field on the image trajectory $x^i(s)$. 
By construction the variation vector field $w^i(s)$ belongs 
to the class of virtual velocities allowed by the constraints.
The variation of the velocity vector $d_w v^\mu$ is to be determined 
by  H\"older's variational principle (\ref{2.10}). We assume
that the variation (\ref{3.2}) of the unconstrained action
($\mu$ is to be replaced by the Cartesian index $i$)
is equal to the H\"older variation (\ref{2.10}) when the former
is restricted on the constrained surface
\begin{equation}
\delta_w S = \int ds (w,[L])\ .
\label{4.0}
\end{equation}
Integrating by parts in the right-hand side of (\ref{4.0}) and
making use of the boundary condition (\ref{3.1}) we
obtain 
\begin{equation}
\left(d_w v, \frac {\partial L}{\partial v}\right) = 
\left(d_vw,\frac{\partial L}{\partial v}\right)\ .
\label{4.1}
\end{equation}  
The integration has been omitted 
because (\ref{4.0}) holds for an arbitrary variation vector field
allowed by the constraints.
On the constraint surface, we have 
$\partial_{v^i}L = \varepsilon_i^\mu \partial_{v^\mu}L,\ 
 \varepsilon_i^\mu= \delta_{ij}g^{\mu\nu}\varepsilon^j_\nu$
and $w^i = \varepsilon^i_\mu w^\mu$.
Relation (\ref{4.1}) leads to
\begin{equation}
(d_w v - d_v w, \varepsilon^\mu) = 0\ ,
\label{4.1a}
\end{equation}
since $\partial_{v^\mu}L$ is also arbitrary.
From the compatibility condition (\ref{2.5}) follows that
\begin{equation}
(d_w v, \varepsilon^\mu)= D_w v^\mu\ ,
\label{4.2}
\end{equation}
and, similarly, 
\begin{equation}
(d_v w, \varepsilon^\mu) = D_v w^\mu\ .
\label{4.3}
\end{equation}
Eq. (\ref{4.1a}) leads to the sought condition that specifies
the variation of the velocity $v^\mu$
\begin{equation}
D_wv^\mu = D_vw^\mu\ .
\label{4.4}
\end{equation}
It can be written as
\begin{equation}
{\cal L}_w v^\mu = d_wv^\mu - d_vw^\mu = 2S^\mu{}_{\sigma\nu}v^\nu w^\sigma\ .
\label{4.5}
\end{equation}
The derivative $d_wv^\mu$ is proportional to
the difference of the vector field $v^\mu$ at two 
neighboring points $q^\mu$ and $q^\mu +u w^\mu, 
u \rightarrow 0$, i.e., $ud_wv^\mu = (v^\mu +ud_wv^\mu)-v^\mu$.
A similar interpretation
holds for $d_vw^\mu$.
The left-hand side of Eq. (\ref{4.5}) 
contains four vectors that can be combined to form
a parallelogram $[w^\mu + (v^\mu +ud_wv^\mu)]
- [v^\mu +(w^\mu +ud_vw^\mu)]$
which is not closed as follows from (\ref{4.5}).
Thus, H\"older's variational principle has led us to the conclusion
that the velocity variation must 
be chosen so that the {\em closure failure} of 
the parallelogram formed by the velocity
and variation vectors would be proportional to the torsion. 
Note also that the closure failure of the 
parallelogram induced by the parallel transport of any two vector
fields along one another (\ref{4.4}) is also
used {\em to define} the torsion on a manifold \cite{schouten}.

Let us take a Lagrangian $L=L(v,q)$ on $TM$ and find the equation
of motion resulting from the new variational principle.
Substituting $d_wv^\mu$ into (\ref{3.2}) and making use
of the boundary condition (\ref{3.1}) to integrate by parts, we get
\begin{equation}
\delta_w S = \int ds w^\mu \left([L]_\mu + 2S^\nu{}_{\mu\sigma}
\frac{\partial L}{\partial v^\nu}v^\sigma\right)\ .
\label{4.6}
\end{equation}
The modified Euler-Lagrange equations are
\begin{equation}
\frac{d}{ds}\frac{\partial L}{\partial v^\mu} -\frac{\partial
  L}{\partial q^\mu} - 2S^\nu{}_{\mu\sigma}\frac{\partial L}{\partial
  v^\nu}
v^\sigma = 0\ .
\label{4.7}
\end{equation}
Eq. (\ref{4.7}) has also been considered in \cite{fk,kp} to describe
the motion in spaces with torsion.

Taking the Lagrangian $L=(v,v)/2$ or $L= -\sqrt{(v,v)}$ in the
auxiliary Euclidean space and restricting it on the constraint surface
$v^i =\varepsilon^i_\mu v^\mu$, we obtain the action
(\ref{1.3}) or (\ref{1.4}). By construction 
the variational principle
$\delta_w S=0$, where the variation on $TM$ is determined by
(\ref{4.4}), should yield the autoparallel equation
(\ref{1.1}). It is not hard to be convinced that the modified
Euler-Lagrange equation (\ref{4.7}) indeed leads to (\ref{1.1}).

A few remarks are in order. Our derivation of the condition
(\ref{4.4}) does not rely on that whether the unconstrained 
Lagrangian explicitly depends on the auxiliary Cartesian
variables $x^i$ or not. The terms involving $\partial_{x^i}L$ 
in Eq. (\ref{4.0}) are cancelled. For this reason the 
final condition (\ref{4.4}) does not depend on the form
of the auxiliary unconstrained Lagrangian and may be applied
to any Lagrangian on the physical configuration space.

The condition (\ref{4.4}) is covariant under general coordinate
transformations on $M$ and has  a transparent geometrical meaning.
A {\em covariant} variation of $v^\mu$ along $w^\mu$ is equal
to a covariant variation of $w^\mu$ along $v^\mu$. All the
modification of (\ref{3.3a}) we have made is that the Lie 
variation has been replaced by the covariant variation.
It is quite remarkable that for tensors on a manifold
there exists only {\em two} independent variations that 
produce tensors out of tensors and involve a displacement 
\cite{schouten}, p.336:
The Lie variation ${\cal L}_w$ and the covariant variation 
$D_w$. Thus, two geometrically distinguished curves on
a manifold, geodesics and autoparallels, can be associated
with the two independent variations available on the manifold
equipped with a connection compatible with the metric. 

If the torsion is not zero, the
variations induced by operators $D_w$ and $D_v$ are
non-coordinate (or anholonomic) because the basis $D_\mu$ in $T_qM$ is
a non-coordinate basis \cite{hehl2}. Indeed, 
assume that there exists a coordinate 
net $q^\mu = q^\mu(s,u)$ such that 
the relation $\partial_{s,u} = 
D_{v,w}$ holds. Taking $F$ to be a scalar, from (\ref{2.7a}) we derive
\begin{equation}
[D_\mu ,D_\nu]F = -2S^\sigma{}_{\mu\nu}D_\sigma F\ .
\label{4.8}
\end{equation}
The curvature term does not contribute to the commutator (\ref{4.8})
because $F$ is a scalar. Thus, $[\partial_s,\partial_u]\neq 0$,
and there is no coordinate system for which the covariant 
derivatives $D_{v,w}$ play the role of the translation operators
along the coordinate lines.

The use of a non-coordinate basis in the tangent space to determine
a variation of the action is not something unusual in mechanics.
Some analogy can be made with Poincar\'e's variational principle
in non-inertial reference frames (e.g. a rigid body in the body-fixed 
reference frame) \cite{poincare,arnold}.
One looks for a {\em re-formulation}
of the Hamilton variational principle in a {\em non-coordinate} basis
in the tangent space of $M$. The coordinate basis in $T_qM$
can always be chosen as $\partial_\mu$ so that a variation
of $F(q)$ is $\delta_\mu F \sim \partial_\mu F$. We can also
assume another basis $e_{\mu'} =e_{\mu'}^\mu(q)\partial_\mu$ 
in the tangent space to determine a variation of any quantity on $M$.
In general, this basis is non-commutative and, hence, non-coordinate
\begin{equation}
[e_{\mu'},e_{\nu'} ] = 2C^{\sigma'}{}_{\mu'\nu'}e_{\sigma'}\ ,
\label{4.9}
\end{equation}  
where the structure functions of the Lie algebra (\ref{4.9})
are coefficients of the object of anholonomity \cite{schouten}.
In  the new basis we have $v^\mu = e^{\mu}_{\mu'} v^{\mu'}$,
$e^{\mu}_{\mu'}e^{\mu'}_{\nu} =\delta_\nu^\mu$. The components
$v^{\mu'}$ of the velocity vector field in a non-coordinate basis 
are called quasivelocities because there is no
$q^{\mu'}(q)$ such that $v^{\mu'} =\dot{q}^{\mu'}$. The problem
is to find a variational principle for the Lagrangian where
the velocity components are taken in the non-coordinate basis,
$\tilde{L}(v^{\mu'}, q^{\mu}) = L(v^\mu,q^{\mu})$. 
It is solved by rewriting Eqs. (\ref{3.2}) and (\ref{3.3})
in the non-coordinate basis. Eq. (\ref{3.3}) assumes the form
(\ref{4.4}) where $\mu$ is replaced by $\mu'$ and the covariant
derivative is taken relative to the connection,
$\Gamma^{\sigma'}{}_{\mu'\nu'}=e^{\sigma'}_\sigma e^{\mu}_{\mu'}
\partial_\mu e^\sigma_{\nu'}$, induced by going
over to the new basis. Such a connection has an antisymmetric
part equal to the object of anholonomity.  
The variational principle $\delta S/\delta w^{\mu'}=0$ 
yields celebrated Poincar\'e's equations. They have the form
(\ref{4.7}) where $L\rightarrow \tilde{L}, \partial_{v^\mu}
\rightarrow \partial_{v^{\mu'}}, \partial_{q^\mu}
\rightarrow e_{\mu'}^\mu\partial_{q^\mu}$ and
$S^\sigma{}_{\mu\nu} \rightarrow C^{\sigma'}{}_{\mu'\nu'}$.

As we see Poincar\'e's variational principle is based on
a {\em coordinate} variation (a smooth deformation of the path), 
but the variation of the velocity components is taken in a non-coordinate
basis in the tangent space. Thus, no torsion force can be gained
by considering a variational principle in an anholonomically
transformed basis. 
By definition \cite{schouten}, the torsion transforms as
a tensor $S^{\sigma'}{}_{\mu'\nu'} = e^{\sigma'}_{\sigma}
 e^{\mu}_{\mu'} e^{\nu}_{\nu'}S^{\sigma}{}_{\mu\nu}$ 
$=(\Gamma^{\sigma'}{}_{\mu'\nu'} -\Gamma^{\sigma'}{}_{\nu'\mu'})/2
-C^{\sigma'}{}_{\mu'\nu'}$. If the torsion is zero in
one basis it is zero in any other. 
The covariant variational principle always 
induces the torsion force
because the condition (\ref{4.4}) is covariant under 
all (coordinate and non-coordinate) transformations of the 
basis in the tangent space. But it should be kept in mind
that the variation specified by the condition (\ref{4.4})
is no smooth deformation of the path with fixed ends.

As a final remark, we shall point out that anholonomic
variations in analytical mechanics 
have also been introduced by Sedov \cite{sedov}
to study dynamics of dissipative systems (they are
examples of non-Lagrangian systems). He also 
proposed an anholonomic variational principle for
such mechanical systems and considered its applications.

\subsubsection{Noether's theorem}
\setcounter{equation}0

In Lagrangian mechanics first integrals of motion
can be obtained from Noether's theorem.
The covariant variational principle has led us to
the new equations of motion (\ref{4.7}). The presence of
the torsion force should affect Noether's integrals of
motion. Therefore it is natural to expect Noether's theorem 
to be modified. 

Consider a one-parameter group of diffeomorphism on a manifold.
Given a trajectory $q^\mu(s)$,
a vector field $\omega^\mu$ determines  a smooth deformation
of the trajectory under the one-parameter group of diffeomorphism on
the manifold
\begin{equation}
d_\omega q^\mu = \omega^\mu\ ,\ \ \
d_\omega v^\mu = \dot{\omega}^\mu\ .
\label{nt1}
\end{equation}
Let the Lagrangian be invariant under the transformations (\ref{nt1})
up to a total time derivative
\begin{equation}
d_\omega L = \frac{d\Phi}{ds}\ ,\ \ \ \Phi = \Phi(v,q,s)\ .
\label{nt2}
\end{equation}
If the motion is determined by the Euler-Lagrange equation,
then in accordance with Noether's theorem \cite{arnold}, the system
possesses the integral of motion
\begin{equation}
\frac{dI}{ds}=0\ ,\ \ \ I= \frac{\partial L}{\partial v^\mu}\,
\omega^\mu -\Phi\ .
\label{nt3}
\end{equation}
The proof follows from relation (\ref{nt2})
that should be written in the form
\begin{equation}
\frac{dI}{ds} + [L]_\mu \omega^\mu = 0\ .
\label{nt4}
\end{equation}
For the actual motion $[L]_\mu =0$, thus leading to the
conservation law (\ref{nt3}). 
Similarly, expressing $[L]_\mu$ from  (\ref{4.7})
and substituting it 
into (\ref{nt4}), we derive modified Noether's theorem
\begin{equation}
\frac{dI}{ds} = 2S^\nu{}_{\mu\sigma}\, \frac{\partial L}{\partial v^\nu}\,
v^\sigma\omega^\mu\ .
\label{nt5}
\end{equation}
In particular, for the time translation symmetry we have
$\omega^\mu = v^\mu$ and $\Phi=L$. The right-hand side of (\ref{nt5})
vanishes since the torsion tensor is antisymmetric. The corresponding
integral of motion is the system energy. On the other hand,
assuming the Lagrangian to be invariant under spatial
translations and rotations (e.g., $L = \frac 12 v^2, g_{\mu\nu}
=\delta_{\mu\nu}$), we observe that the momentum and the
angular momentum are no longer integrals of motion for
a generic torsion tensor.

One can give two additional equivalent formulations of Noether's
theorem. Assume that under the transformations
(\ref{nt1}) the following relation holds
\begin{equation}
d_\omega L = \frac{d\Phi}{ds} - 2S^\nu{}_{\mu\sigma}\,
\frac{\partial L}{\partial v^\nu}\, v^\sigma\omega^\mu\ .
\label{ntm}
\end{equation}
Then (\ref{nt3}) is an integral of motion. 
Clearly, to achieve
(\ref{ntm}), the vector field $\omega^\mu$ should, in general, depend
on the torsion. Although the torsion force violates the 
Noether conservation law as follows from (\ref{nt5}), 
it may also admit new torsion-dependent integrals of motion
\footnote{This conclusion seems to be opposite to the one
made in \cite{kp}.}. To illustrate our statement, consider
two-dimensional motion in the constant metric and torsion 
fields, $S^\mu{}_{\nu\sigma} = \gamma^\mu T_{\nu\sigma}$,
where
$T_{\nu\sigma}=- T_{\sigma\nu}$ is the generator of SO(2), 
$T_{12} =1$, and $ \partial_\mu\gamma^\nu =0, g_{\mu\nu}
=\delta_{\mu\nu}$. The Lagrangian $L=v_\mu^2/2$
exhibits the translational symmetry, but this symmetry 
does not lead to the conservation of the momentum components
as one might see from (\ref{nt5}). 
Nonetheless, we may solve Eq. (\ref{ntm}) relative $\omega^\mu$
and find new integrals instead of the Noether's integrals.
The solution is $\Phi =0, \omega^\mu = [\exp(\varphi T)]^\mu_\nu a^\nu$,
where $a^\mu$ is an arbitrary constant vector
and $\varphi = 2\delta_{\mu\nu}\gamma^\mu q^\mu$. Since $a^\mu$ is
arbitrary, we have two independent integrals of motion
\begin{equation}
I_1 = v_1\cos\varphi -v_2 \sin\varphi\ ,\ \ \ \
I_2 = v_1\sin\varphi + v_2\cos\varphi\ . 
\label{2d}
\end{equation}
The integrals (\ref{2d})
comprise two independent first integrals for the two-dimensional
autoparallel motion in the homogeneous metric and torsion fields.
We remark also that $I_1^2 + I_2^2 = 2E$ where $E=L$ is the energy.

Instead of modifying the transformation law of the Lagrangian,
one can modify the transformation law of its arguments, the generalized
coordinates and velocities. Set
\begin{equation}
d_\omega q^\mu \equiv \omega^\mu\ ,\ \ \
d_\omega v^\mu \equiv d_v\omega^\mu
+ 2S^\mu{}_{\nu\sigma} \omega^\nu v^\sigma\ .
\label{ntm2}
\end{equation}
If the Lagrangian is invariant under these transformations up to a
total time derivative (see (\ref{nt2})), then (\ref{nt3}) is
an integral of motion. The attention should be paid to the fact
that in contrast to the conventional formulation of Noether's
theorem \cite{arnold},
the transformation law (\ref{ntm2}) determines no smooth 
deformation
of the path $q^\mu(s)$ on a manifold
and, in this sense, is not induced
by any diffeomorphism on $M$. In fact, the relations
(\ref{ntm2}) is a postulate that specifies the transformation
law of {\em two independent} coordinates of the tangent bundle.
One first calculates $d_\omega L(v,q)$, then one sets $v^\mu = \dot{q}^\mu$
and looks for such $\omega^\mu = \omega^\mu(v,q)$ that (\ref{nt2})
holds for some $\Phi$. This procedure, though being somewhat unusual,
may turn out to be useful in seeking the integrals of motion
for the modified Euler-Lagrange equations (\ref{4.9}). 

As an example, consider the autoparallel motion in the teleparallel
space (ze\-ro Rie\-mann-Cartan curvature and non-zero torsion). Such
spaces are used to describe a crystal with topological defects
\cite{kroener}. In the anholonomic system discussed in Section 2,
we set $f^i_{\mu\nu}=0$, then the curvature vanishes, while the
torsion does not. We reduce either of the actions (\ref{1.3}) 
and (\ref{1.4}) on the constraint surface and obtain $L=L(v,q)$.
Now it is not hard to verify that the transformations
(\ref{ntm2}) with $w^\mu = (\varepsilon^\mu,a)$, $a^i$ arbitrary
constants, leaves the Lagrangian invariant ($\Phi=0$). Since $a^i$
are arbitrary, the quantities 
\begin{equation}
I^i = \varepsilon^i_\mu(q) v^\mu
\label{ntm3}
\end{equation}
are the integrals of motion. 
Thus, the autoparallel motion in the teleparallel spaces 
can be characterized by a simple property: The velocity 
components taken in the non-coordinate basis ($e_i=\varepsilon^\mu_i
\partial_\mu$ in $T_qM$)  that is 
transported parallel ($D_\mu\varepsilon^i_\nu =0$) 
are the integrals of motion ($\dot{I}^i=0$).
This is, in general, not the case
for non-teleparallel spaces.  

\subsubsection{A Lagrangian formalism for autoparallels}
\setcounter{equation}0

As has been pointed out in the introductionary remarks,
there might exist a non-local action whose
extrema (relative to the conventional variational principle)
would contain autoparallels. In fact, there are infinitely
many such actions. We shall give one of the possible actions.
It has a few additional properties that seems to us
useful for the canonical quantization of the  
autoparallel motion. We require that the sought action should
coincide with a local action whose extrema are geodesics when
the torsion is zero, and the non-locality can be removed
by adding new degrees of freedom coupled to the original 
variables.
Thus, the autoparallel motion
can be modeled by a holonomic dynamics with some auxiliary 
degrees of freedom which
admits both a Hamiltonian formalism and the canonical quantization.

We skip the details of our derivation and just give the answer.
Let $S_g$ be a local action whose extrema are geodesics, i.e.,
we set
\begin{equation}
\frac{\delta S_g}{\delta q^\mu} = g_{\mu\nu}\bar{D}_v v^\nu \ .
\label{da1}
\end{equation}   
Let us introduce an integral operator $\hat{\Lambda}$ by the relation
\begin{equation}
\int ds'{\Lambda}_{\mu\nu}(s,s')\,\frac{\delta}{\delta q^\sigma(s'')}\,
D_v v^\nu(s') = g_{\mu\sigma} \delta(s-s'')\ .
\label{da2}
\end{equation}
Here and below the integration is extended from the initial
to final moment of time.
Consider the non-local action
\begin{equation}
 S = S_g + \int ds\left(
D_vv^\mu -\bar{D}_vv^\mu
\right)\hat{\Lambda}_{\mu\nu} D_vv^\nu
= S_g + \int ds K^{\nu}{}_{\mu\lambda}v^\mu v^\lambda
\hat{\Lambda}_{\nu\sigma} D_vv^\sigma \ .
\label{da3}
\end{equation}
Making use of the properties (\ref{da1}) and (\ref{da2}),
one can convince oneself that from $D_vv^\mu=0$ follows
$\delta_{q^\mu}S = 0$. So the autoparallels are
extrema of the action (\ref{da3}). If the torsion 
is zero, the non-local term in (\ref{da3}) vanishes
and $S=S_g$.

The non-locality of the action (\ref{da3}) still prevents
us from developing a Hamiltonian formalism. We
need a local action. Fortunately, the action (\ref{da3}) can be regarded
as an effective action for physical degrees of freedom in 
a larger dynamical system which is described
by a {\em local} action.  Let us extend the original
configuration space of the system by auxiliary variables
$y^\mu$ whose dynamics is determined by the equation 
\begin{equation}
y^\mu = g^{\mu\nu}\hat{\Lambda}_{\nu\sigma} D_vv^\sigma\ .
\label{da4}
\end{equation}
Consider the action
\begin{equation}
S = S_g + \int ds\left[ K_{\mu\nu\sigma}v^\nu v^\sigma y^\mu
+ \lambda_\mu\left(g^{\mu\nu} \hat{\Omega}_{\nu\sigma} y^\sigma
-D_vv^\mu\right)\right]\ .
\label{da5}
\end{equation}
The operator $\hat{\Omega}$ is to be chosen so that the 
action (\ref{da5}) turns into the action (\ref{da3}) when
it is reduced on the solution of the equation of motion
for the variable $\lambda_\mu$. 

We have
\begin{equation}
g_{\mu\nu}(s)\,\frac{\delta S}{\delta \lambda_\nu(s)}\, =
\int ds' \Omega_{\mu\sigma}(s,s')y^\sigma(s') -g_{\mu\nu}D_v v^\nu(s) =0\ .
\label{6.5a}
\end{equation}
From Eq. (\ref{da2})
follows that the kernel of the operator $\hat{\Omega}$
can be taken in the form
\begin{equation}
\Omega_{\mu\sigma}(s,s'') = g_{\mu\nu}(s)\,
\frac{\delta}{\delta q^\sigma(s'')}\, D_vv^\nu(s)\ .
\label{6.5b}
\end{equation}
With this choice Eq. (\ref{6.5a}) becomes an ordinary 
linear differential equation of second order for $y^\mu$:
\begin{equation}
\ddot{y}^\mu + \left(\Gamma^\mu{}_{\nu\sigma}
  +\Gamma^\mu{}_{\sigma\nu}\right)v^\sigma \dot{y}^\nu
+y^\nu\partial_\nu\Gamma^\mu{}_{\sigma\lambda}v^\sigma v^\lambda
= D_v v^\mu\ .
\label{6.5c}
\end{equation} 
Its solution is a sum of a general solution to the 
homogeneous equation $\hat{\Omega}_{\mu\nu}y^\nu =0$
and a special solution $y^\mu = g^{\mu\nu}\hat{\Lambda}_{\nu\sigma}
D_vv^\sigma$. To recover the action (\ref{da3}), we have to supplement
Eq. (\ref{6.5c}) by the initial conditions such that the homogeneous
equation has only a trivial solution. This is always possible
thanks to the linearity of the equation. In particular, one
can take $\dot{y}^\mu =y^\mu=0$ at the initial moment of time. 

The local action (\ref{da5}) {\em linearly} depends on the second-order
derivatives $\ddot{y}^\mu$ and $\ddot{q}^\mu$.
Assuming zero boundary conditions
for the variable $\lambda_\mu$, we may remove the second-order
time derivatives by
integrating by parts, thus producing  
the final local Lagrangian that depends
only on the coordinates and velocities in the extended
configuration space and involves no
higher-order time derivatives. 
It is important to observe that the Hessian of the Lagrangian
is {\em not} degenerate, that is, the system exhibits no
constraints. The dynamics of the auxiliary degrees of freedom
$y^\mu$ and $\lambda_\mu$
has been chosen so that when the torsion is present, the coupling
between them and the original variables $q^\mu$  
causes the deviation of the trajectory $q^\mu(s)$ 
from the geodesic, making it the autoparallel.

In the path integral formulation of quantum mechanics, the 
auxiliary degrees of freedom $\lambda_\mu$ and $y^\mu$
must be integrated out to obtain an effective path integral
for the original system. This resembles the Feynman-Vernon
approach to quantum dissipative systems \cite{fv}, where the non-potential
friction force is generated by a special coupling of the oscillator
bath to the system. In our case, the non-potential (non-Lagrangian)
torsion force is modeled by a special coupling to the 
$\lambda$- and $y$-degrees of freedom. Extending this analogy
further, one may expect that quantum mechanics in spaces with
torsion, that favors autoparallels in the classical limit,
can not, in general, be described by the wave function
formalism, rather only the density matrix can be constructed
like for dissipative systems. The nice property which could 
be expected is that in the semiclassical approximation
the transition amplitude generated by the effective action
(\ref{da3}) is given by the geodesic action $S_g$ taken
on the autoparallel, the non-local term of (\ref{da3})
will contribute only to quantum fluctuations because
$D_vv^\mu =0$. The canonical quantization of the model
will be considered elsewhere.  
  
\subsubsection{Conclusions}
\setcounter{equation}0

We have analyzed the autoparallel motion from
the point of view of analytical mechanics and
succeeded to represent it as a very special
anholonomic constrained system. Invoking the
variational principles for anholonomic dynamical
systems, we have established the covariant
variational principle for the autoparallels.
We have also analyzed a modification of Noether's
theorem due to the torsion force. Finally, we
have found a possible local action whose
extrema determined by the conventional
Hamilton variational principle contain the autoparallels for some
degrees of freedom. The model can be canonically quantized
by means of the Dirac formalism \cite{dirac}.
 
\subsubsection{Acknowledgments}
A. Chervyakov and H. Kleinert are thanked for useful
discussions.

\end{document}